\documentstyle[12pt]{article}
\title{The N-body Problem in Tetrad Gravity: a First Step towards the Unified
Description of the Four Interactions.}
\author{\large Luca Lusanna \\[3mm]
\em Sezione INFN di Firenze, \\
\em Largo E.Fermi 2, 50125 Firenze, Italy\\
\em E-mail; lusanna@fi.infn.it}
\date{September 29, 1998}

\begin{document}
\maketitle

\begin{abstract}

After a review of the canonical reduction to the rest-frame Wigner-covariant
instant form of standard theories in Minkowski spacetime, a new formulation of 
tetrad gravity is introduced. Its canonical reduction, also in presence of N 
scalar particles, is done. The modification of the ADM formulation to solve the
deparametrization problem of general relativity (how to recover the rest-frame
instant form for G=0), is presented.

Expanded version of the talks given at the XI Int.Conf. ``Problems of Quantum
Field Theory", Dubna 1998, at the III W.Fairbank Meeting and I ICRA
Network Workshop ``The Lense-Thirring Effect", Roma-Pescara 1998, and at the
13th SIGRAV National Congress, Bari 1998.

\end{abstract}

\section{Electromagnetic, Weak and Strong Interactions in Minkowski
Spacetime.}

The attempt to arrive at a unified 
description of the four interactions [with the matter being either 
Grassmann-valued Dirac fields or relativistic particles] based on 
Dirac-Bergmann theory of constraints, which is needed for the Hamiltonian
formulation of both gauge theories and general relativity, motivated the study
of the canonical reduction of a new formulation of tetrad gravity 
\cite{russo1,russo2,russo3,russo4}. This requires to look at 
general relativity from the canonical point of view
generalizing to it all the results already obtained in the canonical
study of gauge theories, since neither a complete reduction
of gravity with an identification of the physical canonical degrees of
freedom of the gravitational field nor a detailed study of its Hamiltonian
group of gauge transformations (whose infinitesimal generators are the first
class constraints) has ever been pushed till the end in an explicit way.

The research program aiming to express the special relativistic strong, weak and
electromagnetic interactions in terms of Dirac's observables \cite{dirac}
is in an advanced  stage of development\cite{re}. This program is based on
the Shanmugadhasan canonical transformations \cite{sha}: if a system has 1st
class constraints at the Hamiltonian level, then, at least locally, one can
find a canonical basis with as many new momenta as 1st class constraints 
(Abelianization of 1st class constraints), with their conjugate canonical 
variables as Abelianized gauge variables and with the remaining pairs of 
canonical variables as pairs of canonically conjugate Dirac's observables 
(canonical basis of physical variables adapted to the chosen Abelianization;
they give a trivialization of the BRST construction of observables). Putting 
equal to zero the Abelianized gauge variables one defines a local gauge of the
model. If a system with constraints admits one (or more) global
Shanmugadhasan canonical transformations, one obtains one (or more) privileged 
global gauges in which the physical Dirac observables are globally defined and
globally separated from the gauge degrees of freedom [for
systems with a compact configuration space this is impossible]. These
privileged gauges (when they exist) can be called generalized Coulomb gauges.
Second class constraints, when present, are also taken into account by the
Shanmugadhasan canonical transformation\cite{sha}.

Firstly, inspired by Ref.\cite{dira},
the canonical reduction to noncovariant 
generalized Coulomb gauges, with the determination of the physical Hamiltonian
as a function of a canonical basis of Dirac's observables, has been achieved for
the following isolated systems (for them one asks that the 10 conserved 
generators of the Poincar\'e algebra are finite so to be able to use group 
theory; theories with external fields can only be recovered as limits in some
parameter of a subsystem of the isolated system): 

a) Yang-Mills theory with Grassmann-valued
fermion fields \cite{lusa} in the case of a trivial principal
bundle over a fixed-$x^o$ $R^3$ slice of Minkowski spacetime with suitable
Hamiltonian-oriented boundary conditions; this excludes monopole solutions and,
since $R^3$ is not compactified, one has only winding number and no instanton
number. After a discussion of the
Hamiltonian formulation of Yang-Mills theory, of its group of gauge
transformations and of the Gribov ambiguity, the theory has been studied in
suitable  weighted Sobolev spaces where the Gribov ambiguity is absent.
The global Dirac observables are the transverse quantities ${\vec A}_{a\perp}
(\vec x,x^o)$, ${\vec E}_{a\perp}(\vec x,x^o)$ and fermion fields dressed
with Yang-Mills (gluonic) clouds. The nonlocal and nonpolynomial (due to the
presence of classical Wilson lines along flat 
geodesics) physical Hamiltonian has been obtained: it is nonlocal but without 
any kind of singularities, it has the correct Abelian limit if the structure 
constants are turned off, and it contains the explicit realization of the 
abstract Mitter-Viallet metric.

b) The Abelian and non-Abelian SU(2)
Higgs models with fermion fields\cite{lv}, where the
symplectic decoupling is a refinement of the concept of unitary gauge.
There is an ambiguity in the solutions of the Gauss law constraints, which
reflects the existence of disjoint sectors of solutions of the Euler-Lagrange
equations of Higgs models. The physical Hamiltonian and Lagrangian of  the
Higgs phase have been found; the self-energy turns out to be local and
contains a local four-fermion interaction. 

c) The standard SU(3)xSU(2)xU(1) model of elementary particles\cite{lv3}
with \hfill\break Grassmann-valued fermion fields.
The final reduced Hamiltonian contains nonlocal self-energies for the
electromagnetic and color interactions, but ``local ones" for the weak 
interactions implying the nonperturbative emergence of 4-fermions interactions.

The next problem is how to covariantize these results valid in Minkowski
spacetime with Cartesian coordinates. Again the starting point 
was given by Dirac\cite{dirac} with his reformulation of classical field theory 
on spacelike hypersurfaces foliating Minkowski spacetime $M^4$ [the foliation 
is defined by an embedding $R\times \Sigma \rightarrow M^4$, $(\tau ,\vec 
\sigma ) \mapsto z^{(\mu )}(\tau ,\vec \sigma )\in \Sigma_{\tau}$, with 
$\Sigma$ an abstract 3-surface diffeomorphic to $R^3$, with $\Sigma_{\tau}$
its copy embedded in $M^4$ labelled by the value $\tau$ (the Minkowski flat 
indices are $(\mu )$; the scalar parameter $\tau$ labels
the leaves of the foliation, $\vec \sigma$ are curvilinear coordinates on
$\Sigma_{\tau}$ and $(\tau ,\vec \sigma )$ are $\Sigma_{\tau}$-adapted 
holonomic coordinates for $M^4$); this is the classical basis of
Tomonaga-Schwinger quantum field theory]. In this way one gets a parametrized 
field theory with a covariant 3+1 splitting of Minkowski spacetime and
already in a form suited to the coupling to general relativity in its ADM
canonical formulation (see also Ref.\cite{kuchar}
, where a theoretical study of this problem is done in curved spacetimes).
The price is that one has to add as new configuration variables  the points 
$z^{(\mu )}(\tau ,\vec \sigma )$ of the spacelike hypersurface $\Sigma_{\tau}$ 
[the only ones carrying Lorentz indices] and then to define the fields on
$\Sigma_{\tau}$ so that they know  the hypersurface $\Sigma_{\tau}$ of 
$\tau$-simultaneity [for a Klein-Gordon field $\phi (x)$, this new field is
$\tilde \phi (\tau ,\vec \sigma )=\phi (z(\tau ,\vec \sigma ))$: it contains 
the nonlocal information about the embedding]. 
Besides a Lorentz-scalar form of the constraints of the given system, 
from the Lagrangian rewritten on the hypersurface [function of  $z^{(\mu )}$
through the induced metric $g_{ AB}=z^{(\mu )}_{A}\eta
_{(\mu )(\nu )}z^{(\nu )}_{B}$, $z^{(\mu )}_{A}=\partial 
z^{(\mu )}/\partial \sigma^{A}$, $\sigma^{A}=(\tau ,\sigma
^{r})$] one gets four further first class constraints ${\cal H}
_{(\mu )}(\tau ,\vec \sigma )=\rho_{(\mu )}(\tau ,\vec \sigma )-l_{(\mu )}(\tau 
,\vec \sigma )T_{sys}^{\tau\tau}(\tau ,\vec \sigma )-z_{r 
(\mu )}(\tau ,\vec \sigma )T_{sys}^{\tau r}(\tau ,\vec \sigma ) 
\approx 0$ [where $T_{sys}^{\tau\tau}(\tau ,\vec \sigma )$, $T_{sys}
^{\tau r}(\tau ,\vec \sigma )$, are the components of 
the energy-momentum tensor of the system in the holonomic coordinate system  
corresponding to the energy- and momentum-density of the
isolated system; one has $\lbrace {\cal H}_{(\mu )}(\tau ,\vec \sigma ),
{\cal H}_{(\nu )}(\tau ,{\vec \sigma}^{'}) \rbrace =0$]
implying the independence of the description from the choice of the
foliation with spacelike hypersufaces. 
The evolution vector is given by $z^{(\mu )}_{\tau}=N_{[z](flat)}l^{(\mu )}+
N^{r}_{[z](flat)}z^{(\mu )}_{r}$, 
where $l^{(\mu )}(\tau ,\vec \sigma )$ is 
the normal to $\Sigma_{\tau}$ in $z^{(\mu )}(\tau ,\vec \sigma )$ and
$N_{[z](flat)}(\tau ,\vec \sigma )=\sqrt{{}^4g_{\tau\tau}+{}^3g^{rs}\, 
{}^4g_{\tau r}\, {}^4g_{\tau s}}=\sqrt{{}^4g
/{}^3\gamma}$, $N_{[z](flat) r}(\tau ,\vec \sigma )={}^3g
_{rs}(\tau ,\vec \sigma )N^{s}_{[z](flat)}(\tau ,\vec 
\sigma )={}^4g_{\tau r}$, are the flat lapse and shift functions
defined through the metric like in general relativity; however, they are not 
independent variables but functionals of $z^{(\mu )}(\tau ,\vec \sigma )$ in 
Minkowski spacetime.

The original Dirac Hamiltonian contains a piece given by
$\int d^3\sigma \lambda^{(\mu )}(\tau ,\vec \sigma ){\cal H}_{(\mu )}
(\tau ,\vec \sigma )$ with $\lambda^{(\mu )}(\tau ,\vec \sigma )$ are
Dirac multipliers. By using ${}^4\eta^{(\mu )(\nu )}=[l^{(\mu )}l^{(\nu )}-$
$z^{(\mu )}_{r}\, {}^3g^{rs} z^{(\nu )}_{s}](\tau ,
\vec \sigma )$, we can write $\lambda_{(\mu )}(\tau ,\vec 
\sigma ){\cal H}^{(\mu )}(\tau ,\vec \sigma )=[(\lambda_{(\mu )}l^{(\mu )})
(l_{(\nu )}{\cal H}^{(\nu )})-(\lambda_{(\mu )}z^{(\mu )}_{r})({}^3g
^{rs} z_{s (\nu )}{\cal H}^{(\nu )})](\tau ,\vec \sigma )$
${\buildrel {def} \over =}\,
N_{(flat)}(\tau ,\vec \sigma ) (l_{(\mu )}{\cal H}^{(\mu )})(\tau ,\vec \sigma )
$$-N_{(flat) r}(\tau ,\vec \sigma ) ({}^3g^{rs} z_{ s (\nu )}
{\cal H}^{(\nu )})(\tau ,\vec \sigma )$ with the (nonholonomic form of
the) constraints $(l_{(\mu )}{\cal H}^{(\mu )})(\tau ,\vec \sigma )\approx 0$,
$({}^3g^{rs} z_{s (\mu )} {\cal H}^{(\mu )})(\tau ,\vec 
\sigma )\approx 0$, satisfying the universal Dirac algebra .
In this way we have defined new  flat lapse and shift functions 

\begin{eqnarray}
N_{(flat)}(\tau ,\vec \sigma )&=& \lambda_{(\mu )}(\tau ,\vec \sigma ) 
l^{(\mu )}(\tau ,\vec \sigma ),\nonumber \\
N_{(flat) r}(\tau ,\vec \sigma )&=& \lambda_{(\mu )}(\tau ,\vec \sigma )
z^{(\mu )}_{r}(\tau ,\vec \sigma ).
\label{I1}
\end{eqnarray}

\noindent which have the same content of the arbitrary Dirac multipliers
$\lambda_{(\mu )}(\tau ,\vec \sigma )$, namely they multiply primary
first class constraints satisfying the Dirac algebra. In 
\hfill\break Minkowski spacetime
they are quite distinct from the previous lapse and shift functions 
$N_{[z](flat)}$, $N_{[z](flat) r}$, defined starting from the metric. 
Instead in general relativity the lapse and shift functions
defined starting from the 4-metric are also the coefficient (in the canonical
part of the Hamiltonian) of secondary first class constraints satisfying the
Dirac algebra.

In special relativity, it is convenient to restrict ourselves to arbitrary 
spacelike hyperplanes $z^{(\mu )}(\tau ,\vec \sigma )=x^{(\mu )}_s(\tau )+
b^{(\mu )}_{r}(\tau ) \sigma^{r}$. Since they are described by 
only 10 variables [an origin $x^{(\mu )}_s(\tau 
)$ and, on it, three orthogonal spacelike unit vectors generating the fixed 
constant timelike unit normal to the hyperplane], we remain only with 10 first 
class constraints determining the 10 variables conjugate to the hyperplane 
[they are a 4-momentum $p^{(\mu )}_s$ and the six independent degrees of 
freedom hidden in a spin tensor $S^{(\mu )(\nu )}_s$; with these 20 canonical 
variables it is possible to build 10 Poincar\'e generators 
$p^{(\mu )}_s$, ${J}^{(\mu )(\nu )}_s=x^{(\mu )}_sp^{(\nu )}_s-x^{(\nu )}
_sp^{(\mu )}_s+S^{(\mu )(\nu )}_s$] in terms of the variables of the system:
${\tilde {\cal H}}^{(\mu )}(\tau )=p^{(\mu )}_s-p^{(\mu )}
_{(sys)}\approx 0$, ${\tilde {\cal H}}^{(\mu )(\nu )}(\tau )=S^{(\mu )(\nu )}
_s-S^{(\mu )(\nu )}_{(sys)}\approx 0$.

After the restriction to spacelike hyperplanes this piece of Dirac Hamiltonian
is reduced to ${\tilde \lambda}^{(\mu )}(\tau ){\tilde {\cal H}}_{(\mu )}(\tau )
-{1\over 2}{\tilde \lambda}^{(\mu )(\nu )}(\tau ){\tilde {\cal H}}_{(\mu 
)(\nu )}(\tau )$. Since at this stage we have $z^{(\mu )}_{r}(\tau 
,\vec \sigma )\approx b^{(\mu )}_{r}(\tau )$, so that $z^{(\mu )}
_{\tau}(\tau ,\vec \sigma )\approx N_{[z](flat)}(\tau ,\vec \sigma )l
^{(\mu )}(\tau ,\vec \sigma )+N^{r}_{[z](flat)}(\tau ,\vec \sigma )$
\hfill\break
$b^{(\mu )}_{r}(\tau ,\vec 
\sigma )\approx {\dot x}^{(\mu )}_s(\tau )+{\dot b}^{(\mu )}_{r}(\tau )
\sigma^{r}=-{\tilde \lambda}^{(\mu )}(\tau )-{\tilde
\lambda}^{(\mu )(\nu )}(\tau )b_{r (\nu )}(\tau )\sigma^{r}$,
it is only now that we get the coincidence of the two definitions of flat
lapse and shift functions:

\begin{eqnarray} 
N_{[z](flat)}(\tau ,\vec \sigma )&\approx& N_{(flat)}(\tau ,\vec \sigma )=
-{\tilde \lambda}
_{(\mu )}(\tau )l^{(\mu )}-l^{(\mu )}{\tilde \lambda}_{(\mu )(\nu )}(\tau )b
^{(\nu )}_{s}(\tau ) \sigma^{s},\nonumber \\
N_{[z](flat)r}(\tau ,\vec \sigma )&\approx&
N_{(flat )}(\tau ,\vec \sigma )=-{\tilde \lambda}
_{(\mu )}(\tau )b^{(\mu )}_{r}(\tau )-b^{(\mu )}_{r}(\tau ){\tilde
\lambda}_{(\mu )(\nu )}(\tau ) b^{(\nu )}_{s}(\tau ) \sigma^{s}.
\label{I2}
\end{eqnarray}

The 20 variables for the phase space description of a hyperplane are

\noindent
i) $x^{(\mu )}_s(\tau ), p^{(\mu )}_s$, parametrizing the origin of the family
of spacelike hyperplanes. The four constraints ${\cal H}^{(\mu )}(\tau ,\vec 
\sigma )\approx 0$  say that: \hfill\break
a) $p^2_s \approx M^2_{(sys)}$ [$M_{(sys)}$ is the invariant mass of the system]
; \hfill\break
b) $u^{(\mu )}_s(p_s)=p^{(\mu )}
_s/p^2_s \approx [the\, orientation\, of\, the\, four-momentum\, $\hfill\break
$of\, the\, 
isolated\, system\, with\, respect\, to\, an\, arbitrary\, given\, 
external\, observer]$. \hfill\break
The origin $x^{(\mu )}_s(\tau )$ is playing the role of a kinematical
center of mass for the isolated system and may be interpreted as a decoupled 
observer with his parametrized clock.\hfill\break
ii)The are other six independent pairs of degrees of freedom are contained in  
$b^{(\mu )}_A(\tau )$ (with $b^{(\mu )}_{\tau}=l^{(\mu )}$ 
$\tau$-independent and normal to the hyperplanes) and $S^{(\mu )(\nu )}_s=
-S^{(\nu )(\mu )}_s$ with the orthonormality constraints $b^{(\mu )}_A\, 
{}^4\eta_{(\mu )(\nu )} b^{(\nu )}_B={}^4\eta_{AB}$.

However, for each configuration of an isolated system there is a privileged
family of hyperplanes (the Wigner hyperplanes orthogonal to $p^{(\mu )}_s$,
existing when it is timelike) corresponding to the intrinsic rest-frame 
of the isolated system. 
To get this result, we must boost at rest all the 
variables with Lorentz indices by using the standard Wigner boost $L^{(\mu )}{}
_{(\nu )}(p_s,{\buildrel \circ \over p}_s)$ for timelike Poincar\'e orbits, and
then add the gauge-fixings $b^{(\mu )}_{r}(\tau )-
L^{(\mu )}{}_{r}(p_s,
{\buildrel \circ \over p}_s)\approx 0$. Since these gauge-fixings depend on 
$p^{(\mu )}_s$, the final canonical variables, apart $p^{(\mu )}_s$ itself, 
are of 3 types: i) there is a non-covariant center-of-mass variable 
${\tilde x}^{(\mu )}
(\tau )$ [it is only covariant under the little group of timelike Poincar\'e
orbits like the Newton-Wigner position operator]; ii) all
the 3-vector variables become Wigner spin 1 3-vectors [boosts in $M^4$ induce
Wigner rotations on them]; iii) all the other variables are Lorentz scalars.

One obtains 
in this way a new kind of instant form of the dynamics (see Ref.\cite{dira2}), 
the  ``Wigner-covariant 1-time rest-frame instant form"\cite{lus1} with a 
universal breaking of Lorentz covariance. 
It is the special relativistic generalization of
the nonrelativistic separation of the center of mass from the relative motion
[$H={{ {\vec P}^2}\over {2M}}+H_{rel}$]. The role of the center of mass is 
taken by the Wigner hyperplane, identified by the point ${\tilde x}^{(\mu )}
(\tau )$ and by its normal $p^{(\mu )}_s$.

The only surviving four 1st class constraints can be put in the 
following form: i) the vanishing of the total (Wigner spin 1) 3-momentum of the
system $\vec p_{sys}\approx 0$ [with  ${\vec p}
_{sys}=[intrinsic\, three-momentum\, of\, the\, isolated\, system\, inside\, 
the\,$\hfill\break
$ Wigner\, hyperplane]$]: instead of putting a restriction on 
$u^{(\mu )}_s(p_s)$ they say that 
the Wigner hyperplane $\Sigma_{W\, \tau}$ is the intrinsic rest frame
[instead, ${\vec p}_s$ is left arbitrary, since $p^{\mu}_s$ depends
upon the orientation of
the Wigner hyperplane with respect to arbitrary reference frames in $M^4$]; 
ii) $\pm \sqrt{p^2_s}-M_{sys}\approx 0$, saying that the
invariant mass $M_{sys}$ of the system replaces the nonrelativistic  Hamiltonian
$H_{rel}$ for the relative degrees of freedom, after the addition of the
gauge-fixing $T_s-\tau \approx 0$ [identifying the time parameter $\tau$,
labelling the leaves of the foliation,  with 
the Lorentz scalar time of the center of mass in the rest frame,
$T_s=p_s\cdot {\tilde x}_s/M_{sys}$; $M_{sys}$  generates the
evolution in this time]. 

In this special gauge 3 degrees of freedom of the isolated system [a 
3-center-of-mass (Wigner spin 1) variable ${\vec x}_{sys}$ defined inside the 
Wigner hyperplane and conjugate to ${\vec p}_{(sys)}$] become gauge variables 
[the natural gauge fixing for ${\vec p}_{(sys)}\approx 0$ is ${\vec x}_{(sys)}
\approx 0$, so that it coincides with 
the origin $\vec \sigma =0$ of the Wigner hyperplane], while ${\tilde x}
^{(\mu )}$  describes a physical decoupled observer. For N free particles
\cite{lus1} one has ${\vec x}_{(sys)}={\vec \eta}_{+}(\tau )=\sum_{i=1}^N
{\vec \eta}_i(\tau )$.After the gauge-fixing ${\vec \eta}_{+}(\tau )\approx 
0$ we remain only 
with Newtonian-like degrees of freedom with rotational covariance: i) a 
3-coordinate (not Lorentz covariant) ${\vec z}_s=\sqrt{p_s^2}({\vec {\tilde x}}
_s-{{{\vec p}_s}\over {p_s^o}}{\tilde x}^o)$ and its conjugate momentum 
${\vec k}_s={\vec p}_s/\sqrt{p^2_s}$  for the  decoupled
center of mass in $M^4$; ii) a set of relative
conjugate pairs of variables with Wigner covariance inside the Wigner hyperplane
. When fields are present, one needs to find a rest-frame canonical basis of 
center-of-mass and relative variables for fields (in analogy to particles)
to identify ${\vec x}_{(sys)}$. Such a basis has already been
found for a real Klein-Gordon field\cite{lon} and it is under reformulation
on spacelike hypersurfaces \cite{mate}.

The isolated systems till now analyzed to get their rest-frame 
Wigner-covariant generalized
Coulomb gauges [i.e. the subset of global Shanmugadhasan canonical bases, 
which, for each Poincar\'e stratum, are also adapted to the geometry of the
corresponding Poincar\'e orbits with their little groups; these special bases
can be named Poincar\'e-Shanmugadhasan bases for the given Poincar\'e stratum
of the presymplectic constraint manifold (every stratum requires an independent
canonical reduction); till now only the main stratum with
$P^2$ timelike and $W^2\not= 0$ has been investigated] are:

a) The system of N scalar particles with Grassmann electric charges
plus the electromagnetic field \cite{lus1}. The starting configuration 
variables are a 3-vector ${\vec \eta}_i(\tau )$ for each particle [$x^{(\mu )}
_i(\tau )=z^{(\mu )}(\tau ,{\vec \eta}
_i(\tau ))$] and the electromagnetic gauge potentials 
$A_{A}(\tau ,\vec \sigma )={{\partial z^{(\mu )}(\tau ,\vec \sigma )}
\over {\partial \sigma^{A}}} A_{(\mu )}(z(\tau ,\vec \sigma ))$, 
which know  the embedding of
$\Sigma_{\tau}$ into $M^4$. One has to choose the sign of the energy of each
particle, because there are not mass-shell constraints (like $p_i^2-m^2_i\approx
0$) among the constraints of this formulation, due to the fact that one has only
three degrees of freedom for particle, determining the intersection of a 
timelike trajectory and of the spacelike hypersurface $\Sigma_{\tau}$. For
each choice of the sign of the energy of the N particles, one describes only one
of the branches of the mass spectrum of the manifestly covariant approach based
on the coordinates $x^{(\mu )}_i(\tau )$, $p^{(\mu )}_i(\tau )$, 
i=1,..,N, and on
the constraints $p^2_i-m^2_i\approx 0$ (in the free case). In this way, one 
gets a description of relativistic particles with a given sign of the energy
with consistent couplings to fields and valid independently from the quantum
effect of pair production [in the manifestly covariant approach, containing
all possible branches of the particle mass spectrum, the classical counterpart 
of pair production is the intersection of different branches deformed by the
presence of fields]. The final Dirac's observables are: i) the transverse 
radiation field variables ${\vec A}_{\perp}$, ${\vec E}_{\perp}$; 
ii) the particle
canonical variables ${\vec \eta}_i(\tau )$, ${{\vec \kappa}}_i(\tau )$,
dressed with a Coulomb cloud. The physical Hamiltonian contains the Coulomb 
potentials extracted from field theory and there is a regularization of the
Coulomb self-energies due to the Grassmann character of the electric charges
$Q_i$ [$Q^2_i=0$]. The no-radiation conditions ${\vec A}_{\perp}={\vec E}
_{\perp}=0$ identify an approximate (delay is neglected) 
canonical subspace containing only physical charged
particles with mutual instantaneous Coulomb potentials.
In Ref.\cite{lus2} there is the study of the 
Lienard-Wiechert potentials and of Abraham-Lorentz-Dirac equations in this
rest-frame Coulomb gauge and also scalar electrodynamics is reformulated in it.
Also the rest-frame 1-time relativistic statistical mechanics has been developed
\cite{lus1}.

b) The system of N scalar particles with Grassmann-valued color charges plus 
the color SU(3) Yang-Mills field\cite{lus3}: 
it gives the pseudoclassical description of the
relativistic scalar-quark model, deduced from the classical QCD Lagrangian and 
with the color field present. The physical invariant mass of the system is
given in terms of the Dirac observables. From the reduced Hamilton equations  
the second order equations of motion both for the reduced transverse color 
field and the particles are extracted. Then, one studies  the N=2 
(meson) case. A special form of the requirement of having only color singlets, 
suited for a field-independent quark model, produces a ``pseudoclassical 
asymptotic freedom" and a regularization of the quark self-energy. With these
results one can covariantize the bosonic part of the standard model given in
Ref.\cite{lv3}.
 
c) The system of N spinning particles of definite energy [$({1\over 2},0)$ or
$(0,{1\over 2})$ representation of SL(2,C)] with Grassmann electric charges 
plus the electromagnetic field\cite{biga} and that of a Grassmann-valued
Dirac field plus the electromagnetic field (the pseudoclassical basis of QED) 
\cite{bigaz}. In both cases there are geometrical complications connected with 
the spacetime description of the path of electric currents and not only of their
spin structure, suggesting a reinterpretation of the supersymmetric scalar 
multiplet as a spin fibration with the Dirac field in the fiber and the
Klein-Gordon field in the base; a new canonical decomposition of the 
Klein-Gordon field into center-of-mass and relative variables \cite{lon,mate} 
will be helpful to clarify these problems. After their solution and after having
obtained the description of Grassmann-valued chiral fields [this will require
the transcription of the front form of the dynamics in the instant one for the
Poincar\'e strata with $P^2=0$] the rest-frame form of the full standard 
$SU(3)\times SU(2)\times U(1)$ model can be achieved.

All these new pieces of information  will allow, after quantization of this new
consistent relativistic mechanics without the classical problems connected
with pair production, to find the  asymptotic states of the covariant
Tomonaga-Schwinger formulation of quantum field theory on spacelike
hypersurfaces: these states are needed for the theory of quantum bound states
[since Fock states do not constitute a Cauchy problem for the field equations,
because an in (or out) particle can be in the absolute future of another one due
to the tensor product nature of these asymptotic states, bound state equations
like the Bethe-Salpeter one have spurious solutions which are excitations in
relative energies, the variables conjugate to relative times (which are gauge
variables\cite{lus1})]. Moreover, it will be possible to include bound states 
among the asymptotic states.

As said in Ref.\cite{lus2,lus3}, the quantization of these rest-frame
models has to overcome two problems. On the particle
side, the complication is the quantization of the square roots associated
with the relativistic kinetic energy terms: in the free case this has been done
in Ref.\cite{lam} [see Refs.\cite{sqroot} for the complications induced by the
Coulomb potential]. On the field side (all physical
Hamiltonian are nonlocal and, with the exception of the Abelian case,
nonpolynomial, but quadratic in the momenta), the obstacle
is the absence (notwithstanding there is no  no-go theorem) of a complete
regularization and renormalization procedure of electrodynamics (to start with) 
in the Coulomb gauge: see Ref.\cite{cou} (and its bibliography)
for the existing results for QED.

However, as shown in Refs.\cite{lus1,lusa}, the rest-frame instant 
form of dynamics automatically gives a physical ultraviolet cutoff in the 
spirit of Dirac and Yukawa: it is the M$\o$ller radius\cite{mol} 
$\rho =\sqrt{-W^2}c/P^2=|\vec S|c/\sqrt{P^2}$ ($W^2=-P^2{\vec 
S}^2$ is the Pauli-Lubanski Casimir when $P^2 > 0$), namely the classical 
intrinsic radius of the worldtube, around the covariant noncanonical 
Fokker-Pryce center of inertia $Y^{\mu}$, 
inside which the noncovariance of the canonical center of mass ${\tilde
x}^{\mu}$ is concentrated. At the quantum level $\rho$ becomes the Compton 
wavelength of the isolated system multiplied its spin eigenvalue $\sqrt{s(s+1)}$
, $\rho \mapsto \hat \rho = \sqrt{s(s+1)} \hbar /M=\sqrt{s(s+1)} \lambda_M$ 
with $M=\sqrt{P^2}$ the invariant mass and $\lambda_M=\hbar /M$ its Compton
wavelength. Therefore, the criticism to classical relativistic physics, based
on quantum pair production, concerns the testing of distances where, due to the
Lorentz signature of spacetime, one has intrinsic classical covariance problems:
it is impossible to localize the canonical center of mass ${\tilde x}^{\mu}$
adapted to the first class constraints of the system
(also named Pryce center of mass and having the same covariance of the 
Newton-Wigner position operator) in a frame independent way.

Let us remember \cite{lus1}
that $\rho$ is also a remnant in flat Minkowski spacetime of 
the energy conditions of general relativity: since the M$\o$ller
noncanonical, noncovariant center of energy has its noncovariance localized
inside the same worldtube with radius $\rho$ (it was discovered in this way)
\cite{mol}, it turns out that for an extended relativistic system with the
material radius smaller of its intrinsic radius $\rho$ one has: i) its 
peripheral rotation velocity can exceed the velocity of light; ii) its 
classical energy density cannot be positive definite everywhere in every frame. 

Now, the real relevant point is that this ultraviolet cutoff determined by
$\rho$ exists also in Einstein's
general relativity (which is not power counting renormalizable) in the case of
asymptotically flat spacetimes, taking into account the Poincar\'e Casimirs of
its asymptotic ADM Poincar\'e charges (when supertranslations are eliminated 
with suitable boundary conditions; let us remark that Einstein and Wheeler
use closed universes because they don't want to introduce boundary conditions,
but in this way they loose Poincar\'e charges and the possibility to make 
contact  with particle physics and to define spin). The generalization of the 
worldtube of radius $\rho$ to asymptotically flat general relativity with 
matter, could be connected with the unproved cosmic censorship hypothesis.

Moreover, the extended Heisenberg relations  of string theory\cite{ven}, i.e.
$\triangle x ={{\hbar}\over {\triangle p}}+{{\triangle p}\over {T_{cs}}}=
{{\hbar}\over {\triangle p}}+{{\hbar \triangle p}\over {L^2_{cs}}}$ implying the
lower bound $\triangle x > L_{cs}=\sqrt{\hbar /T_{cs}}$ due to the $y+1/y$
structure,
have a counterpart in the quantization of the M$\o$ller radius\cite{lus1}:
if we ask that, also at the quantum level, one cannot test the inside of the 
worldtube, we must ask $\triangle x > \hat \rho$ which is the lower bound
implied by the modified uncertainty relation $\triangle x ={{\hbar}\over 
{\triangle p}}+{{\hbar \triangle p}\over {{\hat \rho}^2}}$. This could imply 
that the center-of-mass canonical noncovariant  3-coordinate 
$\vec z=\sqrt{P^2}({\vec {\tilde x}}-{{\vec P}\over {P^o}}{\tilde x}^o)$ 
\cite{lus1} cannot become a
self-adjoint operator. See Hegerfeldt's theorems (quoted in 
Refs.\cite{lusa,lus1}) and his interpretation 
pointing at the impossibility of a good localization of relativistic particles
(experimentally one determines only a worldtube in spacetime emerging from the 
interaction region). Since the eigenfunctions of the canonical center-of-mass
operator are playing the role of the wave function of the universe, one could 
also say that the center-of-mass variable has not to be quantized, because it
lies on the classical macroscopic side of Copenhagen's interpretation and,
moreover, because, in the spirit of Mach's principle that only relative 
motions can be observed, no one can observe it (it is only used to define a
decoupled ``point particle clock"). On the other hand, if one 
rejects the canonical noncovariant center of mass in favor of the covariant
noncanonical Fokker-Pryce center of inertia $Y^{\mu}$, $\{ Y^{\mu},Y^{\nu} \}
\not= 0$, one could invoke the philosophy of quantum groups to quantize 
$Y^{\mu}$ to get some kind of quantum plane for the center-of-mass 
description. Let us remark that the quantization of the square root Hamiltonian
done in Ref.\cite{lam} is consistent with this problematic.

In conclusion, the best set of canonical coordinates adapted to the constraints
and to the geometry of Poincar\'e orbits in Minkowski spacetime
and naturally predisposed to the
coupling to canonical tetrad gravity is emerging for the electromagnetic, weak
and strong interactions with matter described either by fermion fields or by
relativistic particles with a definite sign of the energy.

\section{Canonical Reduction of Tetrad Gravity, the Deparametrization Problem 
and the N-body Problem.}

Tetrad gravity is the formulation of general relativity natural for the
coupling to the fermion fields of the standard model. However, we need a
formulation of it, which allows to solve its constraints for doing the
canonical reduction and to solve
the deparametrization problem of general relativity (how to recover
the rest-frame instant form when the Newton constant is put equal to zero,
G=0).

To implement this program
we shall restrict ourselves to the simplest class of spacetimes 
[time-oriented pseudo-Riemannian or Lorentzian 4-manifold $(M^4,{}^4g)$ with 
signature $\epsilon \, (+---)$ ($\epsilon =\pm 1$) and with a choice of time 
orientation], assumed to be:

i) Globally hyperbolic 4-manifolds, i.e. topologically they are $M^4=R\times 
\Sigma$, so to have a well posed Cauchy problem [with $\Sigma$ the abstract
model of Cauchy surface] at least till when no singularity develops in $M^4$
[see the singularity theorems]. Therefore, these spacetimes admit regular 
foliations with orientable, complete, non-intersecting spacelike 3-manifolds
$\Sigma_{\tau}$ [$\tau :M^4 
\rightarrow R$, $z^{\mu} \mapsto \tau (z^{\mu})$, is a global timelike
future-oriented function labelling the leaves (surfaces of simultaneity)]. In 
this way, one obtains 3+1 splittings of $M^4$ and the possibility of a 
Hamiltonian formulation.

ii) Asymptotically flat at spatial infinity, so to have the possibility to 
define asymptotic Poincar\'e charges \cite{adm,reg,reg1,reg2,reg3,ash}: they 
allow the definition of a M$\o$ller radius also in general relativity and are a
bridge towards a future soldering with the theory of elementary particles in
Minkowski spacetime defined as irreducible representation of its kinematical,
globally implemented Poincar\'e group according to Wigner. 

iii) Admitting a spinor (or spin) structure\cite{wald} for the coupling
to fermion fields. Since we 
consider noncompact space- and time-orientable spacetimes, spinors can be
defined if and only if they are ``parallelizable" \cite{ger}, like in our case.
This implies that the orthonormal frame principal SO(3)-bundle over 
$\Sigma_{\tau}$ (whose connections are the spin connections determined by the 
cotriads) is trivial.

iv) The noncompact parallelizable simultaneity 3-manifolds (the Cauchy surfaces)
$\Sigma_{\tau}$ are assumed to be topologically trivial, geodesically complete
and, finally, diffeomorphic to
$R^3$. These 3-manifolds have the same manifold structure as Euclidean spaces:
a) the geodesic exponential map $Exp_p:T_p\Sigma_{\tau}\rightarrow \Sigma
_{\tau}$ is a diffeomorphism ; b) the sectional curvature is 
less or equal  zero everywhere; c) they have no ``conjugate locus" [i.e.
there are no pairs of conjugate Jacobi points (intersection points of distinct
geodesics through them) on any geodesic] and no ``cut locus" [i.e. no closed
geodesics through any point].

v) Like in Yang-Mills case \cite{lusa}, the 3-spin-connection on the orthogonal
frame SO(3)-bundle (and therefore cotriads) will have to be 
restricted to suited weighted Sobolev spaces to avoid Gribov ambiguities
\cite{lusa,moncr}. In
turn, this implies the absence of isometries of the noncompact Riemannian
3-manifold $(\Sigma_{\tau},{}^3g)$ [see for instance the review paper in Ref.
\cite{cho}]. 

Diffeomorphisms on $\Sigma_{\tau}$ ($Diff\, \Sigma_{\tau}$) are interpreted 
in the passive way, following Ref.\cite{be}, in accord with the Hamiltonian
point of view that infinitesimal diffeomorphisms are generated by taking the
Poisson bracket with the 1st class supermomentum constraints [passive
diffeomorphisms are also named `pseudodiffeomorphisms'].

The new formulation of tetrad gravity [see Refs.
\cite{weyl,dirr,schw,kib,tetr,char,maluf,hen1,hen2,hen3,hen4} for the existing
versions of the theory] utilizes the ADM action of metric gravity with the
4-metric expressed in terms of arbitrary cotetrads. Let us remark that both
in the ADM metric and tetrad formulation one has to introduce the extra
ingredient of the 3+1 splittings of $M^4$ with foliations whose leaves $\Sigma
_{\tau}$ are spacelike 3-hypersurfaces. However, their points $z^{\mu}(\tau
,\vec \sigma )$ [$(\tau ,\vec \sigma )$ are $\Sigma_{\tau}$-adapted holonomic
coordinates of $M^4$] are not configurational variables of these theories in
contrast to what happens in Minkowski parametrized theories [${{\partial 
z^{\mu}}\over {\partial \sigma^A}}$ are not tetrads when $M^4$ is not
Minkowski spacetime with Cartesian coordinates, because ${}^4g^{AB}{{\partial 
z^{\mu}}\over {\partial \sigma^A}} {{\partial z^{\nu}}\over {\partial 
\sigma^B}}={}^4g^{\mu\nu}\not= {}^4\eta^{(\mu )(\nu )}$].

By using $\Sigma_{\tau}$-adapted holonomic coordinates for $M^4$, one has 
found a new \hfill\break
parametrization of arbitrary tetrads and cotetrads on $M^4$ in 
terms of cotriads on $\Sigma_{\tau}$ [${}^3e_{(a)r}(\tau ,\vec \sigma )$], of
lapse [$N(\tau ,\vec \sigma )$] and shift [$N_{(a)}(\tau ,\vec \sigma )=
\{{}^3e_{(a)r} N^r\} (\tau ,\vec \sigma )$] functions and of 3 parameters
[$\varphi_{(a)}(\tau ,\vec \sigma )$] parametrizing point-dependent Wigner 
boosts for timelike Poincar\'e orbits. Putting these variables in the ADM 
action for metric gravity \cite{adm} (with the 3-metric on
$\Sigma_{\tau}$ expressed in terms of cotriads: ${}^3g_{rs}={}^3e_{(a)r}\,
{}^3e_{(a)s}$ with positive signature), one gets a new action
depending only on lapse, shifts and cotriads, but not on the boost parameters
(therefore, there is no need to use Schwinger's time gauge). There are 10
primary and 4 secondary first class constraints and a vanishing canonical
Hamiltonian. Besides the 3 constraints associated with the vanishing
Lorentz boost momenta, there are
4 constraints saying that the momenta associated with lapse and shifts vanish,
3 constraints describing rotations, 3 constraints generating 
space-diffeomorphisms on the cotriads induced by those ($Diff\, \Sigma_{\tau}$)
on $\Sigma_{\tau}$ (a linear combination of supermomentum constraints and
of the rotation ones;a different combination of these constraints generates
SO(3) Gauss law constraints for the momenta ${}^3{\tilde \pi}^r_{(a)}$
conjugated to cotriads with the covariant derivative built with the spin
connection) and one superhamiltonian constraint. It turns out that 
with the technology developed for Yang-Mills theory, one can Abelianize the 3 
rotation constraints and then also the space-diffeomorphism constraints. 
In the Abelianization of the rotation constraints one needs the Green function
of the 3-dimensional covariant derivative containing the spin connection,
well defined only if there is no Gribov ambiguity in the SO(3)-frame bundle 
and no isometry of the Riemannian 3-manifold
$(\Sigma_{\tau},{}^3g)$. The Green function is similar
to the Yang-Mills one for a principal SO(3)-bundle \cite{lusa}, but, instead
of the Dirac distribution for the Green function of the flat
divergence, it contains the DeWitt function or bitensor \cite{dew} defining
the tangent in one endpoint of the geodesic arc connecting two points (which 
reduces to the 
Dirac distribution only locally in normal coordinates). Moreover, the definition
of the Green function now requires the geodesic exponential map.

In the resulting quasi-Shanmugadhasan canonical basis, 
the original cotriad can be expressed in closed
form in terms of 3 rotation angles, 3 diffeomorphism-parameters and a
reduced cotriad depending only on 3 independent variables.
(they are Dirac's observables with respect to 13 of the 14 first class 
constraints) and with their conjugate momenta, still subject to the reduced 
form of the superhamiltonian constrain: this is the phase space over the 
superspace of 3-geometries\cite{witt}.

Till now no coordinate condition\cite{cji} has been imposed. It turns out that
these conditions are hidden in the choice of how to parametrize the reduced
cotriads in terms of three independent functions. The
simplest parametrization (the only one studied till now) corresponds to choose a
system of global 3-orthogonal coordinates on $\Sigma_{\tau}$, in which the 
3-metric is diagonal. With a further canonical transformation on the reduced 
cotriads and conjugate momenta, one arrives at a canonical basis containing the 
conformal factor $\phi (\tau ,\vec \sigma )=e^{q(\tau ,\vec \sigma )/2}$ of the 
3-geometry and its conjugate momentum $\rho (\tau ,\vec \sigma )$ plus two 
other pairs of conjugate canonical variables $r_{\bar a}(\tau ,\vec \sigma )$,
$\pi_{\bar a}(\tau ,\vec \sigma )$, $\bar a = 1,2$. 
The reduced superhamiltonian 
constraint, expressed in terms of these variables, turns out to be an integral 
equation for the conformal factor (reduced Lichnerowicz equation) whose 
conjugate momentum is, therefore, the last gauge variable. If we replace the
gauge fixing of the Lichnerowicz\cite{conf} and York\cite{york,yoyo,ciuf}
approach [namely the vanishing of the trace of the extrinsic curvature of
$\Sigma_{\tau}$, ${}^3K(\tau ,\vec \sigma )\approx 0$, also named the internal
extrinsic York time\cite{qadir}] with the natural one $\rho (\tau ,\vec \sigma )
\approx 0$ and we go to Dirac brackets, we find that $r_{\bar a}(\tau ,\vec 
\sigma )$, $\pi_{\bar a}(\tau ,\vec \sigma )$ are the canonical basis for the 
physical degrees of freedom or Dirac's observables of the gravitational field.

Since we have this physical canonical basis, it is possible to define ``void 
spacetimes" as the equivalence class of spacetimes ``without gravitational 
field", whose members in the 3-orthogonal gauge are obtained 
by adding by hand the second class constraints $r_{\bar a}(\tau ,\vec 
\sigma )\approx 0$, $\pi_{\bar a}(\tau ,\vec \sigma )\approx 0$ [one gets $\phi
(\tau ,\vec \sigma )=1$ as the relevant solution of the reduced Lichnerowicz
equation]. The members of this equivalence class (the extension to general 
relativity of the Galilean non inertial coordinate systems with their Newtonian 
inertial forces) are in general ``not flat" [their $\Sigma_{\tau}$'s are
3-conformally flat] but gauge equivalent to Minkowski spacetime with 
Cartesian coordinates (this holds in absence of matter).

The next step is to find the physical Hamiltonian for them and to solve the
deparametrization problem.
If we wish to arrive at the soldering
of tetrad gravity with matter and parametrized Minkowski formulation for the 
same matter, we must require that the lapse and shift functions of tetrad 
gravity [which must grow linearly in $\vec \sigma$, in suitable asymptotic
Minkowski coordinates, according to the existing literature on asymptotic
Poincar\'e charges at spatial infinity \cite{reg,reg1}] 
must agree asymptotically with the flat lapse and shift functions, which,
however, are unambigously defined only on Minkowski spacelike hyperplanes.

In metric ADM gravity the canonical Hamiltonian is $H_{(c)ADM}=\int d^3\sigma
[N {\tilde {\cal H}}+N_r {\tilde {\cal H}}^r](\tau ,\vec \sigma )\approx 0$,
where ${\tilde {\cal H}}(\tau ,\vec \sigma )\approx 0$ and ${\tilde {\cal H}}
^r(\tau ,\vec \sigma )\approx 0$ are the superhamiltonian and supermomentum 
constraints. It is differentiable and finite only for suitable $N(\tau ,\vec 
\sigma )=n(\tau ,\vec \sigma )\, {\rightarrow}_{|\vec \sigma |\rightarrow 
\infty}\, 0$, $N_r(\tau ,\vec \sigma )=n_r(\tau ,\vec \sigma )\, {\rightarrow}
_{|\vec \sigma |\rightarrow \infty}\, 0$ defined in Ref\cite{reg1} in
suitable asymptotic coordinate systems. For more general lapse and shift
functions one must add a surface term \cite{witt} to $H_{(c)ADM}$, which 
contains the ``strong" Poincar\'e charges \cite{adm} $P^A_{ADM}$,
$J^{AB}_{ADM}$ [they are conserved and 
gauge invariant surface integrals]. To have well defined asymptotic Poincar\'e
charges at spatial infinity\cite{adm,reg,reg1,reg2} one needs: i) the selection
of a class of coordinates systems for $\Sigma_{\tau}$ asymptotic to flat 
coordinates; ii) the choice of a class of Hamiltonian boundary conditions in
these coordinate systems [all the fields must belong to some functional space 
of the type of the weighted Sobolev spaces]; iii) a definition of the 
Hamiltonian group ${\cal G}$ of gauge transformations (and in particular of
proper gauge transformations) with a well defined 
limit at at spatial infinity so to respect i) and ii). The scheme is the same
needed to define the non-Abelian charges in Yang-Mills theory\cite{lusa}. The
delicate point is to be able to exclude supertranslations\cite{wald}, because
the presence of these extra asymptotic charges leads to the replacement of the
asymptotic Poincar\'e group with the infinite-dimensional spi group\cite{ash},
which does not allow the definition of the Poincar\'e spin due to the absence of
the Pauli-Lubanski Casimir. This can be done with suitable boundary conditions
(in particular all the fields and gauge transformations must have
direction independent limits at spatial infinity) respecting the ``parity 
conditions" of Ref.\cite{reg1} [see also Ref.\cite{reg2}].

Let us then remark that
in Ref.\cite{drc} and in the book in Ref.\cite{dirac} (see also Ref.\cite{reg}),
Dirac introduced asymptotic Minkowski rectangular coordinates $z^{(\mu )}
_{(\infty )}(\tau ,\vec \sigma )=x^{(\mu )}_{(\infty )}(\tau )$\hfill\break
$+b^{(\mu )}
_{(\infty )\, r}(\tau ) \sigma^{r}$ in $M^4$ at spatial infinity
$S_{\infty}=\cup_{\tau} S^2_{\tau ,\infty}$ 
For each value of $\tau$, the coordinates $x^{(\mu )}
_{(\infty )}(\tau )$ labels a point, near spatial infinity 
chosen as origin. On it there is a flat tetrad $b^{(\mu )}_{(\infty )\, A}
(\tau )= \{ \, l^{(\mu )}_{(\infty )}=b^{(\mu )}_{(\infty )\, \tau}=\epsilon
^{(\mu )}{}_{(\alpha )(\beta )(\gamma )} b^{(\alpha )}_{(\infty )\, 1}
(\tau )b^{(\beta )}_{(\infty )\, 2}(\tau )b^{(\gamma )}_{(\infty )\, 
3}(\tau );$\hfill\break
$\, b^{(\mu )}_{(\infty )\, r}(\tau )\, \}$, with
$l^{(\mu )}_{(\infty )}$ $\tau$-independent, satisfying $b^{(\mu )}_{(\infty )\,
A}\, {}^4\eta_{(\mu )(\nu )}\, b^{(\nu )}_{(\infty )\, B}={}^4\eta_{AB}$ for
every $\tau$ [at this level we do not assume that $l^{(\mu )}_{(\infty )}$ is
tangent to $S_{\infty}$, as the normal $l^{\mu }$ to $\Sigma_{\tau}$]. 
There will be transformation coefficients $b^{\mu}_A(\tau ,\vec 
\sigma )$ from the holonomic adapted coordinates $\sigma^A=(\tau ,\sigma
^{r})$ to coordinates $x^{\mu}=z^{\mu}(\sigma^A)$ in an atlas of $M^4$,
such that in a chart at spatial infinity one has $z^{\mu}(\tau ,\vec \sigma )
=\delta^{\mu}_{(\mu )} z^{(\mu )}(\tau ,\vec \sigma )$ and $b^{\mu}_A(\tau 
,\vec \sigma )= \delta^{\mu}_{(\mu )} b^{(\mu )}_{(\infty )A}(\tau )$
[for $r\, \rightarrow \, \infty$ one has ${}^4g_{\mu\nu}\, \rightarrow \,
\delta^{(\mu )}_{\mu}\delta^{(\nu )}_{\nu}{}^4\eta_{(\mu )(\nu )}$ and
${}^4g_{AB}=b^{\mu}_A\, {}^4g_{\mu\nu} b^{\nu}_B\, \rightarrow \,
b^{(\mu )}_{(\infty )A}\, {}^4\eta_{(\mu )(\nu )} b^{(\nu )}_{(\infty )B}=
{}^4\eta_{AB}$ ].

Dirac\cite{drc} and, then, Regge and
Teitelboim\cite{reg} proposed that the asymptotic Minkowski rectangular
coordinates $z^{(\mu )}_{(\infty )}(\tau ,\vec \sigma )=x^{(\mu )}_{(\infty )}
(\tau )+b^{(\mu )}_{(\infty ) r}(\tau )\sigma^{r}$ should define 
10 new independent degrees of freedom at the spatial boundary $S_{\infty}$, 
as it happens for Minkowski parametrized theories\cite{lus1} when restricted 
to spacelike hyperplanes [defined by $z^{(\mu )}(\tau ,\vec \sigma )\approx 
x^{(\mu )}_s(\tau )+b^{(\mu )}_{r}(\tau )\sigma^{r}$]; then, 
10 conjugate momenta should exist.
These 20 extra variables of the Dirac proposal
can be put in the form: $x^{(\mu )}_{(\infty )}(\tau )$,
$p^{(\mu )}_{(\infty )}$, $b^{(\mu )}_{(\infty ) A}(\tau )$ [with $b^{(\mu )}
_{(\infty ) \tau }=l^{(\mu )}_{(\infty )}$ $\tau$-independent], $S^{(\mu )(\nu 
)}_{(\infty )}$, with  Dirac brackets implying the orthonormality
constraints $b^{(\mu )}_{(\infty ) A}\, {}^4\eta_{(\mu )(\nu )} b^{(\nu )}
_{(\infty ) B}={}^4\eta_{AB}$ [so that $p^{(\mu )}_{(\infty )}$ and 
$J^{(\mu )(\nu )}_{(\infty )}=x^{(\mu )}_{(\infty )}p^{(\nu )}_{(\infty )}-
x^{(\nu )}_{(\infty )}p^{(\mu )}_{(\infty )}+S^{(\mu )(\nu )}_{(\infty )}$
satisfy a Poincar\'e algebra]. In analogy with Minkowski parametrized
theories restricted to spacelike hypersurfaces, one expects to have 10 extra
first class constraints of the type $p^{(\mu )}_{(\infty )}-P^{(\mu )}
_{ADM}\approx 0$, $S^{(\mu )(\nu )}_{(\infty )}-S^{(\mu )(\nu )}_{ADM}
\approx 0$ with $P^{(\mu )}_{ADM}$, $S^{(\mu )(\nu )}_{ADM}$ related to the
ADM Poincar\'e charges $P^A_{ADM}$, $J^{AB}_{ADM}$.
The origin $x^{(\mu )}_{(\infty )}$ is going to play
the role of a decoupled observer with his parametrized clock.

Let us remark that if we
replace $p^{(\mu )}_{(\infty )}$ and $S^{(\mu )(\nu )}_{(\infty )}$, whose
Poisson algebra is the direct sum of an Abelian algebra of translations and of 
a Lorentz algebra, with the new variables (with holonomic indices with
respect to $\Sigma_{\tau}$) $p^A_{(\infty )}=b^A_{(\infty )(\mu )}p^{(\mu )}
_{(\infty )}$, $J^{AB}_{(\infty )}=b^A_{(\infty )(\mu )}b^B_{(\infty )(\nu )}
S^{(\mu )(\nu )}_{(\infty )}$ [$\not= b^A_{(\infty )(\mu )}b^B_{(\infty )(\nu )}
J^{(\mu )(\nu )}_{(\infty )}$], the Poisson brackets for $p^{(\mu )}
_{(\infty )}$, $b^{(\mu )}_{(\infty ) A}$, $S^{(\mu )(\nu )}_{(\infty )}$ 
imply that $p^A_{(\infty )}$, $J^{AB}_{(\infty )}$ satisfy a Poincar\'e
algebra. This  implies that the Poincar\'e generators ${P}^A_{ADM}$, ${J}^{AB}
_{ADM}$ define in the asymptotic Dirac rectangular coordinates a momentum 
${P}^{(\mu )}_{ADM}$ and only an  ADM spin tensor ${S}^{(\mu )(\nu )}_{ADM}$
[to define an angular momentum tensor $J^{(\mu )(\nu )}_{ADM}$ one should
find a ``center of mass of the gravitational field" $X^{(\mu )}_{ADM} [{}^3g,
{}^3{\tilde \Pi}]$ (see Ref.\cite{lon} for the Klein-Gordon case) conjugate to
$P^{(\mu )}_{ADM}$, so that $J^{(\mu )(\nu )}_{ADM}=X^{(\mu )}_{ADM}P^{(\nu )}
_{ADM}-X^{(\nu )}_{ADM}P^{(\mu )}_{ADM}+S^{(\mu )(\nu )}_{ADM}$].

The following splitting of the lapse and shift functions and the following set
of boundary conditions [consistent with the ones of Ref.\cite{ck}]
fulfill all the previous requirements [soldering with
the lapse and shift functions on Minkowski hyperplanes; absence of 
supertranslations (strictly speaking one gets $P^r_{ADM}=0$ like in Ref.
\cite{ck}; it is an open problem how to weaken the condition on ${}^3{\tilde
\Pi}^{rs}$ without reintroducing supertranslations); $r=|\vec \sigma |$]

\begin{eqnarray}
{}^3g_{rs}(\tau ,\vec \sigma )\, &{\rightarrow}_{r\, 
\rightarrow \infty}&\, (1+{M\over r})
\delta_{rs}+{}^3h_{rs}(\tau 
,\vec \sigma )=(1+{M\over r})
\delta_{rs}+o_4(r^{-3/2}),\nonumber \\
{}^3{\tilde \Pi}^{rs}(\tau ,\vec \sigma )\, &{\rightarrow}
_{r\, \rightarrow \infty}&\, {}^3k^{rs}(\tau ,\vec \sigma )=
o_3(r^{-5/2}),\nonumber \\
&&{}\nonumber \\
N(\tau ,\vec \sigma )&=& N_{(as)}(\tau ,\vec \sigma )
+n(\tau ,\vec \sigma ),\quad\quad n(\tau ,\vec 
\sigma )\, = O(r^{-(3+\epsilon )}),\nonumber \\
N_{r}(\tau ,\vec \sigma )&=&N_{(as)r}(\tau ,\vec \sigma )+
n_{r}(\tau ,\vec \sigma ),\quad\quad 
n_{r}(\tau ,\vec \sigma )\, = O(r^{-\epsilon}),\nonumber \\
&&{}\nonumber \\
N_{(as)}(\tau ,\vec \sigma )&=&-{\tilde \lambda}_{(\mu )}(\tau )l^{(\mu )}
_{(\infty )}-l^{(\mu )}_{(\infty )}{\tilde \lambda}_{(\mu )(\nu )}(\tau )
b^{(\nu )}_{(\infty)s}(\tau ) \sigma^{s}=\nonumber \\
&=&-{\tilde \lambda}_{\tau}(\tau )-{\tilde \lambda}_{\tau s}(\tau )
\sigma^{s},\nonumber \\
N_{(as)r}(\tau ,\vec \sigma )&=&-b^{(\mu )}_{(\infty )r}(\tau )
{\tilde \lambda}_{(\mu )}(\tau )-b^{(\mu )}_{(\infty )r}(\tau ){\tilde 
\lambda}_{(\mu )(\nu )}(\tau ) b^{(\nu )}_{(\infty )s}(\tau ) \sigma
^{s}=\nonumber \\
&=&-{\tilde \lambda}_r(\tau )-{\tilde \lambda}_{rs}(\tau ) \sigma^s,\nonumber \\
\Rightarrow&& N_{(as)A}(\tau ,\vec \sigma )\, {\buildrel {def} \over =}\,
(N_{(as)}\, ;\, N_{(as) r}\, )(\tau ,\vec \sigma )
=-{\tilde \lambda}_A(\tau )-{1\over 2}{\tilde \lambda}_{As}(\tau ) 
\sigma^{s},\nonumber \\
&&{}
\label{III14}
\end{eqnarray}

\noindent with${}^3h_{rs}(\tau ,-\vec \sigma )={}^3h_{rs}(\tau ,\vec \sigma )$,
${}^3k^{rs}(\tau ,-\vec \sigma )=-{}^3k^{rs}(\tau ,\vec \sigma )$; here 
${}^3{\tilde \Pi}^{rs}(\tau ,\vec \sigma )$ is the momentum conjugate to the
3-metric ${}^3g_{rs}(\tau ,\vec \sigma )$ in ADM metric gravity.

After the addition of the surface term, the resulting canonical and Dirac
Hamiltonians of ADM metric gravity are

\begin{eqnarray}
{H}_{(c)ADM}&=&
\int d^3\sigma [(N_{(as)}+n) {\tilde {\cal H}}+(N_{(as)r}+
n_{r})\, {}^3{\tilde {\cal H}}^{r}](\tau ,\vec \sigma )
\mapsto\nonumber \\
&\mapsto& {H}^{'}_{(c)ADM}=
\int d^3\sigma [(N_{(as)}+n) {\tilde {\cal H}}+(N_{(as)r}+
n_{r})\, {}^3{\tilde {\cal H}}^{r}](\tau ,\vec \sigma )+
\nonumber \\
&+&{\tilde \lambda}_A(\tau )
P^A_{ADM}+{\tilde \lambda}_{AB}(\tau ) J^{AB}_{ADM} =\nonumber \\
&=&\int d^3\sigma [ n {\tilde {\cal H}}+n_{r}\, 
{}^3{\tilde {\cal H}}^{r}](\tau ,\vec \sigma )+
{\tilde \lambda}_A(\tau ) {\hat P}^A_{ADM}+{\tilde \lambda}_{AB}(\tau )
{\hat J}^{AB}_{ADM}\approx \nonumber \\
&\approx& {\tilde \lambda}_A(\tau ) {\hat P}^A_{ADM}+{\tilde 
\lambda}_{AB}(\tau ) {\hat J}^{AB}_{ADM},\nonumber \\
{H}^{'}_{(D)ADM}&=&{H}^{'}_{(c)ADM}+\int d^3\sigma [\lambda_n {\tilde
\pi}^n+\lambda^{\vec n}_r {\tilde \pi}^r_{\vec n}](\tau ,\vec \sigma )+
\nonumber \\
&+&\zeta_A(\tau ) {\tilde \pi}^A(\tau )+\zeta_{AB}(\tau ) {\tilde \pi}
^{AB}(\tau ),
\label{III10}
\end{eqnarray}

\noindent with the  ``weak conserved improper charges"
${\hat P}^A_{ADM}$, ${\hat J}_{ADM}^{AB}$ [they are volume integrals differing
from the weak charges by terms proportional to integrals of the constraints].

The previous splitting implies to replace the variables $N(\tau ,\vec \sigma )$,
$N_r(\tau ,\vec \sigma )$ with the ones ${\tilde \lambda}_A(\tau )$, ${\tilde 
\lambda}_{AB}(\tau )=-{\tilde \lambda}_{BA}(\tau )$, $n(\tau ,\vec \sigma )$,
$n_r(\tau ,\vec \sigma )$ [with conjugate momenta ${\tilde \pi}^A(\tau )$, 
${\tilde \pi}^{AB}(\tau )=-{\tilde \pi}^{BA}(\tau )$, ${\tilde \pi}^n(\tau
,\vec \sigma )$, ${\tilde \pi}^r_{\vec n}(\tau ,\vec \sigma )$]
in the ADM theory.

With these assumptions one has the following form of the line element (also its
form in tetrad gravity is given)

\begin{eqnarray}
ds^2&=& 
\epsilon ( [N_{(as)}+n]^2 - [N_{(as)r}+n_{r}] 
{}^3g^{rs}[N_{(as)s}+n_{s}] ) (d\tau )^2-
\nonumber \\
&-&2\epsilon [N_{(as)r}+n_{r}] d\tau d\sigma^{r} -
\epsilon \, {}^3g_{rs} d\sigma^{r} d\sigma^{s}=
\nonumber \\
&=&\epsilon ( [N_{(as)}+n]^2 - [N_{(as) r}+n_r] {}^3e^r_{(a)}\, {}^3e^s
_{(a)} [N_{(as) s}+n_s] ) (d\tau )^2-\nonumber \\
&-&2\epsilon [N_{(as) r}+n_r] d\tau d\sigma^r -\epsilon \, {}^3e_{(a)r}\, 
{}^3e_{(a)s} d\sigma^r d\sigma^s.
\label{I3b}
\end{eqnarray}

The final suggestion of Dirac is to modify ADM metric gravity in the
following way: \hfill\break
i) add the 10 new primary constraints $p^A_{(\infty )}-{\hat P}^A_{ADM}
\approx 0$, $J^{AB}_{(\infty )}-{\hat J}^{AB}_{ADM}\approx 0$, where
$p^A_{(\infty )}=b^A_{(\infty )(\mu )}p^{(\mu )}_{(\infty )}$, $J^{AB}_{(\infty
)}=b^A_{(\infty )(\mu )}b^B_{(\infty )(\nu )}S^{(\mu )(\nu )}_{(\infty )}$
[remember that $p^A_{(\infty )}$ and $J^{AB}_{(\infty )}$ satisfy a Poincar\'e 
algebra];\hfill\break
ii) consider ${\tilde \lambda}_A(\tau )$, ${\tilde \lambda}_{AB}(\tau )$, as
Dirac multipliers  for these 10
new primary constraints, and not as configurational 
(arbitrary gauge) variables coming from the
lapse and shift functions [so that there are no conjugate momenta 
${\tilde \pi}^A(\tau )$, ${\tilde \pi}^{AB}(\tau )$ and no associated Dirac
multipliers $\zeta_A(\tau )$, $\zeta_{AB}(\tau )$], in the
assumed Dirac Hamiltonian [it is finite and differentiable]

\begin{eqnarray}
H_{(D)ADM}&=& \int d^3\sigma [ n {\tilde {\cal H}}+n_r {\tilde {\cal H}}^r
+\lambda_{n} {\tilde \pi}^n+\lambda^{\vec n}_r{\tilde \pi}^r_{\vec n}]
(\tau ,\vec \sigma )-\nonumber \\
&-&{\tilde \lambda}_A(\tau ) [p^A_{(\infty )}-{\hat P}^A
_{ADM}]-{\tilde \lambda}_{AB}(\tau )[J^{AB}_{(\infty )}-
{\hat J}^{AB}_{ADM}]\approx 0,
\label{III16}
\end{eqnarray}

The reduced phase space is still the ADM one: on the ADM variables there are 
only the secondary first class constraints ${\tilde {\cal H}}(\tau ,\vec 
\sigma )\approx 0$, ${\tilde {\cal H}}^r(\tau ,\vec \sigma )\approx 0$  
[generators of proper gauge transformations], because the other
first class constraints $p^A_{(\infty )}-{\hat P}^A_{ADM}\approx 0$, $J^{AB}
_{(\infty )}-{\hat J}^{AB}_{ADM}\approx 0$ do not generate improper gauge
transformations but eliminate 10 of the extra 20 variables.

In this modified ADM metric gravity, one has restricted the 3+1 splittings of 
$M^4$ to foliations whose leaves $\Sigma_{\tau}$ tend to Minkowski spacelike
hyperplanes asymptotically at spatial infinity in a direction independent
way. Therefore, these $\Sigma^{'}_{\tau}$ should be determined by the 10
degrees of freedom $x^{(\mu )}_{(\infty )}(\tau )$, $b^{(\mu )}_{(\infty 
)A}(\tau )$, like it happens for flat spacelike hyperplanes: this means that 
it must be possible to define a ``parallel transport" of the asymptotic tetrads 
$b^{(\mu )}_{(\infty )A}(\tau )$ to get well defined tetrads in each point of 
$\Sigma^{'}_{\tau }$. While it is not yet clear whether this can be done for 
${\tilde \lambda}_{AB}(\tau )\not= 0$ [maybe Nester's  teleparallelism
\cite{nester} may be used], there is a solution for ${\tilde \lambda}
_{AB}(\tau )=0$. This case corresponds
to go to the Wigner-like hypersurfaces [the
analogue of the Minkowski Wigner hyperplanes with the asymptotic normal
$l^{(\mu )}_{(\infty )}=l^{(\mu )}_{(\infty )\Sigma}$ parallel to ${\hat P}
^{(\mu )}_{ADM}$]. Following the same procedure defined for Minkowski 
spacetime, one gets ${\bar S}^{rs}_{(\infty )}\equiv {\hat J}^{rs}_{ADM}$ [see 
Ref.\cite{lus1} for the definition of ${\bar S}^{AB}_{(\infty )}$],
${\tilde \lambda}_{AB}(\tau )=0$ and [$\epsilon_{(\infty )}=
\sqrt{p^2_{(\infty )}}$] $-{\tilde \lambda}_A(\tau ) [p^a_{(\infty )}-
{\hat P}^A_{ADM}] =
-{\tilde \lambda}_{\tau}(\tau ) [\epsilon_{(\infty )}-{\hat P}^{\tau}_{ADM}]
+{\tilde \lambda}_{r}(\tau ) {\hat P}^{r}_{ADM}$, so that the final form of
these four surviving constraints is $\epsilon_{(\infty )}-{\hat P}^{\tau}
_{ADM} \approx 0$, ${\hat P}^{r}_{ADM}\approx 0$.

On this subclass of foliations [whose leaves $\Sigma^{(WSW)}_{\tau}$ will be 
called Wigner-Sen-Witten hypersurfaces] one can introduce a parallel
transport by using the interpretation of Ref.\cite{p26} of the Witten
spinorial method of demonstrating the positivity of the ADM energy \cite{p27}.
Let us consider the Sen-Witten connection \cite{sen,p27} restricted to
$\Sigma_{\tau}^{(WSW)}$ (it depends on the trace of the extrinsic curvature
of $\Sigma_{\tau}^{(WSW)}$) and the spinorial Sen-Witten equation associated
with it. As shown in Ref.\cite{p28}, this spinorial equation can be
rephrased as an equation whose solution determines (in a surface dependent
dynamical way) a tetrad in each point of $\Sigma_{\tau}^{(WSW)}$ once it is 
given at spatial infinity (again this requires a direction independent
limit).

On the Wigner-Sen-Witten hypersurfaces (the intrinsic asymptotic rest frame
of the gravitational field), the remaining four extra constraints 
are: ${\hat P}^{r}_{ADM}\approx 0$ (this is automatically implemented 
with the boundary conditions of Ref.\cite{ck})
and $\epsilon_{(\infty )}=\sqrt{p^2
_{(\infty )}}\approx {\hat P}^{\tau }_{ADM} \approx M_{ADM}=\sqrt{{\hat P}
^2_{AM}}$. Now the spatial indices have become spin-1 Wigner indices [they
transform with Wigner rotations under asymptotic Lorentz transformations].
As said for parametrized theories in Minkowski spacetime, in this special
gauge 3 degrees of freedom of the gravitational field [ a 3-center-of-mass
variable ${\vec X}_{ADM}[{}^3g,{}^3{\tilde \Pi}]$ inside the Wigner-Sen-Witten
hypersurface] become gauge variables, while ${\tilde x}^{(\mu )}_{(\infty )}$
[the canonical non covariant variable replacing $x^{(\mu )}_{(\infty )}$]
becomes a decoupled observer with his ``point particle clock"
\cite{ish,kuchar1} near spatial infinity.

Since the positivity theorems for the ADM 
energy imply that one has only timelike or lightlike
orbits of the asymptotic Poincar\'e group, the restriction to universes with
timelike ADM 4-momentum allows to define the M\"oller radius
$\rho_{AMD}=\sqrt{-{\hat W}^2_{ADM}}/{\hat P}^2_{ADM}c$ from the asymptotic 
Poincar\'e Casimirs ${\hat P}^2_{ADM}$, ${\hat W}^2_{ADM}$ [it is an intrinsic
classical unit of length like in parametrized Minkowski theories, to be used
as an ultraviolet cutoff in a future attempt of quantization]. 

By going from ${\tilde x}^{(\mu )}
_{(\infty )}$, $p^{(\mu )}_{(\infty )}$, to the canonical basis $T
_{(\infty )}=p_{(\infty )(\mu )}{\tilde x}^{(\mu )}_{(\infty )}/\epsilon
_{(\infty )}=p_{(\infty )(\mu )}x^{(\mu )}_{(\infty )}/\epsilon_{(\infty )}
{}{}{}$, $\epsilon_{(\infty )}$, $z^{(i)}_{(\infty )}=\epsilon_{(\infty )}
({\tilde x}^{(i)}_{(\infty )}-p^{(i)}_{(\infty )}{\tilde x}^{(o)}_{(\infty )}
/p^{(o)}_{(\infty )})$, $k^{(i)}_{(\infty )}=p^{(i)}_{(\infty )}/\epsilon
_{(\infty )}=u^{(i)}(p^{(\rho )}_{(\infty )})$, 
like in the flat case one finds that the final 
reduction requires the gauge-fixings $T_{(\infty )}-\tau \approx 0$ and
$X^{r}_{ADM}\approx 0$, where $\sigma^{r}=X^{r}_{ADM}$ is a
variable representing the ``center of mass" of the 3-metric of the slice
$\Sigma_{\tau}$ of the asymptotically flat spacetime $M^4$. 
Since $\{ T_{(\infty )},\epsilon_{(\infty )} \}=-\epsilon$, with the
gauge fixing $T_{(\infty )}-\tau \approx 0$ one gets ${\tilde \lambda}_{\tau}
(\tau )\approx \epsilon$, and the final Dirac 
Hamiltonian is $H_D=M_{ADM}+{\tilde \lambda}_r(\tau ) {\hat P}^r_{ADM}$ with 
$M_{ADM}$ [the ADM mass of the universe] the natural physical Hamiltonian to 
reintroduce an evolution in the ``mathematical" $T_{(\infty )}\equiv
\tau$: namely in the rest-frame time identified with the parameter $\tau$
labelling the leaves $\Sigma_{\tau}^{(WSW)}$ of the foliation of $M^4$. 
Physical times (atomic clocks, ephemeridis time...) must be put in a local
1-1 correspondence with this ``mathematical" time.

The asymptotic rest-frame instant form
realization of the Poincar\'e generators becomes (no more reference to the 
boosts ${\hat J}^{\tau r}_{ADM}$)

\begin{eqnarray}
&&\epsilon_{(\infty )}=M_{ADM},\nonumber \\
&&p^{(i)}_{(\infty )},\nonumber \\
&&J^{(i)(j)}_{(\infty )}={\tilde x}^{(i)}_{(\infty )}p^{(j)}_{(\infty )}-
{\tilde x}^{(j)}_{(\infty )} p^{(i)}_{(\infty )} +\delta^{(i)r}\delta
^{(j)s}{\hat J}^{rs}_{ADM},\nonumber \\
&&J^{(o)(i)}_{(\infty )}=p^{(i)}_{(\infty )} {\tilde x}^{(o)}_{(\infty )}-
\sqrt{M^2_{ADM}+{\vec p}^2_{(\infty )}} {\tilde x}^{(i)}_{(\infty )}-
{ {\delta^{(i)r}{\hat J}^{rs}_{ADM} \delta^{(s(j)} 
p^{(j)}_{(\infty )} }\over
{M_{ADM}+\sqrt{M^2_{ADM}+{\vec p}^2_{(\infty )}} } } .
\label{I24}
\end{eqnarray}

All this construction holds also in our formulation of tetrad gravity 
(since it uses the ADM action) and in
its canonically reduced form in the 3-orthogonal gauge. In
particular the Poincar\'e charges of void spacetimes vanish.

Therefore, the final physical Hamiltonian for the physical gravitational field 
is the reduced volume form of the ADM energy ${\hat P}^{\tau}_{ADM}[r_{\bar a}.
\pi_{\bar a}, \phi (r_{\bar a},\pi_{\bar a})]$ with $\phi$ solution of the
reduced Lichnerowicz equation in the 3-orthogonal gauge with $\rho (\tau
,\vec \sigma )\approx 0$.

Let us compare the standard generally covariant formulation of gravity based on
the Hilbert action with its invariance under $Diff\, M^4$ with the ADM
Hamiltonian formulation.

Regarding the 10 Einstein equations of the standard approach,
the Bianchi identities imply that four equations are linearly dependent on the 
other six ones and their gradients. Moreover, the four combinations of
Einstein's equations projectable to phase space (where they become the
secondary first class superhamitonian and supermomentum constraints of canonical
metric gravity) are independent from the accelerations being restrictions
on the Cauchy data. As a consequence
the Einstein equations have solutions, in which the ten
components ${}^4g_{\mu\nu}$ of the 4-metric depend on only two truly dynamical
degrees of freedom (defining the physical gravitational field) and on eight
undetermined degrees of freedom.
This transition from the ten components ${}^4g
_{\mu\nu}$ of the tensor ${}^4g$ in some atlas of $M^4$ to the 2 
(deterministic)+8 (undetermined) degrees of freedom breaks general covariance,
because these quantities are neither tensors nor invariants under
diffeomorphisms (their functional form is atlas dependent).

Since the Hilbert action is invariant under $Diff\, M^4$, one usually says
that a ``gravitational field" is a 4-geometry over $M^4$, namely an equivalence
class of spacetimes $(M^4, {}^4g)$, solution of Einstein's equations, modulo
$Diff\, M^4$. See, however, the interpretational problems about what is 
observable in general relativity for instance in Refs.\cite{rove,rov}, in
particular the facts that i) scalars under $Diff\, M^4$, like ${}^4R$, are not
Dirac's observables but gauge dependent quantities; ii) the functional form
of ${}^4g_{\mu\nu}$ in terms of the physical gravitational field and, 
therefore, the angle and distance properties of material bodies and the
standard procedures of defining measures of length and time based on the
line element $ds^2$, are gauge dependent.

Instead in the ADM formalism with the extra notion of 3+1 splittings of $M^4$,
the (tetrad) metric ADM action (differing from the Hilbert one by a surface
term) is quasi-invariant under the (14) 8 types of gauge transformations
which are the pull-back of the Hamiltonian group ${\cal G}$ of gauge
transformations, whose generators are the first class constraints of the theory
. The Hamiltonian group ${\cal G}$ has a subgroup (whose generators are the 
supermomentum and superhamiltonian constraints) formed by the diffeomorphisms
of $M^4$ adapted to its 3+1 splittings, $Diff\, M^{3+1}$ [it is different
from $Diff\, M^4$]. Moreover, the Posson algebra of the supermomentum and 
superhamiltonian constraints reflects the embeddability in $M^4$ of the
foliation associated with the 3+1 splitting \cite{tei}. 
Now a ``gravitational field" is the equivalence
class of spacetimes modulo the Hamiltonian group ${\cal G}$, and different 
members of the equivalence class have in general different 4-Riemann tensors
[these equivalence classes are connected with the conformal 3-geometries of the
Lichnerowicz-York approach and contain different gauge-related 4-geometries].

The interpretation of the 14 gauge transformations and of their gauge fixings
in tetrad gravity (it is independent from the presence of matter) is the 
following [a tetrad in a point of $\Sigma_{\tau}$ is a local observer]
:\hfill\break
i) the gauge fixings of the gauge boost parameters associated with the 3 boost
constraints are equivalent to choose the local observer either at rest or
Lorentz-boosted;\hfill\break
ii) the gauge fixings of the gauge angles associated with the 3 rotation
constraints are equivalent to the fixation of the standard of non rotation of
the local observer;\hfill\break
iii) the gauge fixings of the 3 gauge parameters associated with the passive
space diffeomorphisms [$Diff\, \Sigma_{\tau}$; change of coordinates charts]
are equivalent to a fixation of 3 standards of length by means of a choice of
a coordinate system on $\Sigma_{\tau}$ [the measuring apparatus (the ``rods")
should be defined in terms of Dirac's observables for some kind of matter, after
its introduction into the theory];\hfill\break
iv) according to constraint theory the choice of 3-coordinates on $\Sigma
_{\tau}$ induces the gauge fixings of the 3 shift functions [i.e. of ${}^4g
_{oi}$], whose gauge nature is connected with the ``conventionality of
simultaneity" \cite{havas} [therefore, the gauge fixings are equivalent
to a choice of simultaneity and, as a consequence, to a statement about the
isotropy or anisotropy of the velocity of light in that gauge];\hfill\break
v) the gauge fixing on the the momentum $\rho (\tau ,\vec \sigma )$ conjugate
to the conformal factor of the 3-metric [this gauge variable is the source of the
gauge dependence of 4-tensors and of the scalars under $Diff\, M^4$, together
with the gradients of the lapse and shift functions] is a nonlocal statement
about the extrinsic curvature of the leaves $\Sigma_{\tau}$ of the given 3+1
splitting of $M^4$; since the superhamiltonian constraint produces normal
deformations of $\Sigma_{\tau}$ \cite{tei} and, therefore, transforms a 3+1
splitting of $M^4$ into another one (the ADM formulation is independent from
the choice of the 3+1 splitting), this gauge fixing is equivalent to the
choice of a particular 3+1 splitting;\hfill\break
vi) the previous gauge fixing induces the gauge fixing of the lapse function
(which determines the packing of the leaves $\Sigma_{\tau}$ in the chosen
3+1 splitting) and, therefore, is equivalent to the fixation of a standard of
proper time [again ``clocks" should be built with the Dirac's observables
of some kind of matter].

The 3-orthogonal gauge of tetrad gravity is the equivalent of the Coulomb 
gauge in classical electrodynamics (like the harmonic gauge is the equivalent 
of the Lorentz gauge). Only after a complete gauge fixing the 4-tensors and 
the scalars under $Diff\, M^4$ become measurable quantities (like the 
electromagnetic vector potential in the Coulomb gauge).  At this stage it 
becomes acceptable the proposal of Bergmann\cite{be} of identifying the points
of a spacetime $(M^4, {}^4g)$, solution of the Einstein's equations in
absence of matter, in a way invariant under spacetime diffeomorphisms extended 
to 4-tensors (so that the rule is separately valid for each 4-geometry 
contained in the equivalence class of Dirac's observables defining a 
gravitational field), by using four invariants bilinear in the
Weyl tensors called ``individuating fields''(see also Refs.\cite{rove,rov}).

Our approach breaks the general covariance of general relativity completely by
going to the special 3-orthogonal gauge. But this is done in a way
naturally associated with theories with first class
constraints (like all formulations of general relativity and the standard model
of elementary particles with or without supersymmetry): the global
Shanmugadhasan canonical transformations (when they exist) 
correspond to privileged Darboux charts for presymplectic manifolds defined
by the first class constraints. Therefore, the gauges identified by 
these canonical transformations should have a special (till now unexplored) 
role also also in generally covariant theories, in which traditionally one 
looks for observables invariant under diffeomorphisms
 and not for not generally covariant 
Dirac observables.

Let us remember that Bergmann\cite{be} made the following critique of 
general covariance: it would be desirable to restrict the group of
coordinate transformations (spacetime diffeomorphisms) in such a way that it
could contain an invariant subgroup describing the coordinate transformations 
that change the frame of reference of an outside observer (these 
transformations could be called Lorentz transformations; see also the
comments in Ref.\cite{ll} on the asymptotic behaviour of coordinate
transformations); the remaining 
coordinate transformations would be like the gauge transformations of
electromagnetism. This is what we have done. In this way
``preferred' coordinate systems will emerge the WSW hypersurfaces), 
which, as said by Bergmann, are
not ``flat": while the inertial coordinates are determined experimentally by 
the observation of trajectories of force-free bodies, these intrinsic 
coordinates can be determined only by much more elaborate experiments, since
they depend, at least, on the inhomogeneities of the ambient gravitational 
fields.
See also Ref.\cite{ellis} for other critics to general covariance: very often
to get physical results one uses preferred coordinates not merely for
calculational convenience, but also for understanding. In Ref.\cite{elli} this
fact has been formalized as the ``principle of restricted covariance". In our
case the choice of the gauge-fixings has been dictated by the Shanmugadhasan
canonical transformations, which produce generalized Coulomb gauges, in which 
one can put in normal form the Hamilton equations for the canonical
variables of the gravitational field [and, therefore, they also produce a normal
form of the two associated combinations of the Einstein equations which
depend on the accelerations].

If we add to the tetrad ADM action the action for N scalar particles with
positive energy in the form of Ref.\cite{lus1} [where it was given on
arbitrary Minkowski spacelike hypersurfaces], the only constraints which are
modified are the superhamiltonian one, which gets a dependence on the matter 
energy density ${\cal M}(\tau ,\vec \sigma )$, and the 3 space diffeomorphism
ones, which get a dependence on the matter momentum density ${\cal M}_r(\tau
,\vec \sigma )$. The canonical reduction and the determination of the Dirac 
observables can be done like in absence of matter. However, the reduced 
Lichnerowicz equation for the conformal factor of the 3-metric in the 
3-orthogonal gauge and with $\rho (\tau ,\vec \sigma )\approx 0$ acquires now
an extra dependence on ${\cal M}(\tau ,\vec \sigma )$ and ${\cal M}_r(\tau 
,\vec \sigma )$. 

Since, as a preliminary result, we are interested in
identifying explicitly the action-at-a-distance 
(Newton-like and gravitomagnetic) potentials among particles hidden in tetrad
gravity (like the Coulomb potential is hidden in the electromagnetic gauge
potential), we shall restrict ourselves to void spacetimes without
gravitational field by adding the second class constraints $r_{\bar a}
(\tau ,\vec \sigma )\approx 0$, $\pi_{\bar a}(\tau ,\vec \sigma )\approx 0$.
Now void spacetimes are no more gauge equivalent to Minkowski spacetime in
Cartesian coordinates. Moreover, since in presence of matter the equations of 
motion do not imply $r_{\bar a}=\pi_{\bar a}=0$, void spacetimes have now to 
be understood as a strong approximation identifying the instantaneous
action-at-a-distance interaction among the particles contained in tetrad
gravity. If we develop the conformal factor $\phi (\tau ,\vec 
\sigma )$ in a formal series in the Newton constant G [$\phi =1+\sum_{n=1}
^{\infty} G^n \phi_n$], one can find a solution $\phi = 1+G \phi_1$ at order G
(post-Minkowskian approximation) of the reduced Lichnerowicz equation. However, 
due to a self-energy divergence in $\phi$ evaluated at the positions ${\vec
\eta}_i(\tau )$ of the particles, one needs to rescale the bare masses to
physical ones, $m_i\, \mapsto \, \phi^{-2}(\tau ,{\vec \eta}_i(\tau )) 
m_i^{(phys)}$, and to make a regularization of the type defined in Refs.
\cite{ein}. Then, the regularized solution for $\phi$ can be put in the reduced 
form of the ADM energy, which becomes [${\vec \kappa}_i(\tau )$ are the
particle momenta conjugate to ${\vec \eta}_i(\tau )$; ${\vec n}_{ij}=
[{\vec \eta}_i-{\vec \eta}_j]/|{\vec \eta}_i-{\vec \eta}_j|$]

\begin{eqnarray}
{\hat P}^{\tau}_{ADM}&=&\sum_{i=1}^Nc\sqrt{m_i^{(phys) 2}c^2+{\vec \kappa}^2
_i(\tau )}-\nonumber \\
&-&{G\over {c^2}}\sum_{i\not= j} {{ \sqrt{m_i^{(phys) 2}c^2+{\vec 
\kappa}^2_i(\tau )}\, \sqrt{m_j^{(phys) 2}c^2+{\vec \kappa}^2_j(\tau )} }\over
{|{\vec \eta}_i(\tau )-{\vec \eta}_j(\tau )|}}-\nonumber \\
&-&{G\over {8c^2}} \sum_{i\not= j} {{ 3{\vec \kappa}_i(\tau )\cdot {\vec 
\kappa}_j(\tau )-5 {\vec \kappa}_i(\tau )\cdot {\vec n}_{ij}(\tau ) 
{\vec \kappa}_j(\tau )\cdot {\vec n}_{ij}(\tau )}\over
{|{\vec \eta}_i(\tau )-{\vec \eta}_j(\tau )|}} + O(G^2).
\label{1}
\end{eqnarray}

One sees the Newton-like and the gravitomagnetic (in the sense of York)
potentials (both of them need regularization) at the post-Minkowskian level
(order G but exact in c) emerging from the tetrad ADM version of Einstein
general relativity. For G=0 we recover N free scalar particles on the Wigner
hyperplane in Minkowski spacetime, as required by deparametrization. For
$c\, \rightarrow \, \infty$, we get the post-Newtonian Hamiltonian

\begin{eqnarray}
H_{PN}&=&\sum_{i=1}^N{{ {\vec \kappa}_i^2(\tau )}\over {2m_i^{(phys)}}}
(1-{{2G}\over {c^2}}\sum_{j\not= i}{{m_j^{(phys)}}\over 
{|{\vec \eta}_i(\tau )-{\vec \eta}_j(\tau )|}})-{G\over 2}\sum_{i\not= j}
{{m_i^{(phys)}\, m_j^{(phys)}}\over
{|{\vec \eta}_i(\tau )-{\vec \eta}_j(\tau )|}}-\nonumber \\
&-&{G\over {8c^2}}\sum_{i\not= j} {{ 3{\vec \kappa}_i(\tau )\cdot {\vec 
\kappa}_j(\tau )-5 {\vec \kappa}_i(\tau )\cdot {\vec n}_{ij}(\tau ) 
{\vec \kappa}_j(\tau )\cdot {\vec n}_{ij}(\tau )}\over
{|{\vec \eta}_i(\tau )-{\vec \eta}_j(\tau )|}} + O(G^2),
\label{2}
\end{eqnarray}

\noindent which is of the type of the ones implied by the results of
Refs.\cite{ein,droste} [the differences 
are probably connected with the use of different coordinate
systems and with the fact that one has essential singularities on the particle
worldlines and the need of regularization].

The future research program will concentrate on the following subjects:

1)the post-Minkowskian 2-body problem in void spacetimes, to see the relevance 
of exact relativistic recoil effects in the motion of binaries;

2) the replacement of scalar particles with spinning ones to identify the
precessional effects (like the Lense-Thirring one) of gravitomagnetism;

3) the linearization of the theory in the 3-orthogonal gauge in presence of
matter: besides finding the Coulomb gauge description of gravitational
waves, one expects to find a consistent (post-Minkowskisn) coupling of the
linearized gravitational field with matter, since the Bianchi identities have
been solved, and to go beyond the strong approximation of void spacetimes;

4) perfect fluids and, then, extended relativistic bodies;

5) the coupling of  tetrad gravity to the
electromagnetic field, to fermion fields and then to the standard model,
trying to make to reduction to Dirac's observables in all these cases and to
study their post-Minkowskian approximations;

6) quantization of tetrad gravity in the 3-orthogonal gauge with $\rho (\tau 
,\vec \sigma )\approx 0$: for each perturbative (in G) solution of the reduced 
Lichnerowicz equation one defines a Schroedinger equation in $\tau$ for a wave 
functional $\Psi [\tau ; r_{\bar a}]$ with the associated quantized ADM energy
${\hat P}^{\tau}_{ADM}[r_{\bar a}, i{{\delta}\over {\delta r_{\bar a}}} ]$
as Hamiltonian; no problem of physical scalar product is present, but only
ordering problems in the Hamiltonian; moreover, one has the Moller radius as
a ultraviolet cutoff.

\end{document}